\documentclass{amsart}
\usepackage{amssymb, amsmath}
\usepackage{physics}
\usepackage{mathrsfs}
\usepackage{amscd}
\usepackage[active]{srcltx}
\usepackage{verbatim}
\usepackage[colorlinks,linkcolor={blue}, citecolor={blue},urlcolor={red},]{hyperref}

\let\mathcal \undefined
\def\mathcal{\mathscr}

\usepackage{newsymbol}
\let\emptyset \undefined
\let\ge       \undefined
\let\le       \undefined
\newsymbol\le          1336  
\newsymbol\ge          133E  
\newsymbol\emptyset    203F
\newsymbol\notle       230A
\newsymbol\notge       230B

\newcommand{\ov}{\overline}
\newcommand{\F}{{\ov{F}}}
\newcommand{\N}{{\ov{N}}}
\renewcommand{\L}{{\ov{L}}}
\newcommand{\W}{{\ov{W}}}
\newcommand{\up}{\uparrow}
\newcommand{\down}{\downarrow}

\newcommand{\ok}{{\rm ok}}
\newcommand{\ovok}{\ov{\rm ok}}
\newcommand{\fail}{{\rm fail}}
\newcommand{\ovfail}{\ov{\rm fail}}

\allowdisplaybreaks

\begin{document}

 \title[Relational Frauchiger--Renner]{Relational analysis of the Frauchiger--Renner paradox and interaction-free detection of records from the past}

 \author{Marijn Waaijer \& Jan van Neerven}
 \address{Delft University of Technology, Department of Applied Mathematics, P.O. Box 5031, 2600 GA Delft, The Netherlands}
 \email{waaijermarijn@gmail.com, J.M.A.M.vanNeerven@TUDelft.nl}

 \date{\today}


\begin{abstract} We present an analysis of the Frauchiger--Renner Gedankenexperiment from the point
of view of the relational interpretation of quantum mechanics.
Our analysis shows that the paradox obtained by Frauchiger and Renner disappears if one rejects promoting one agent's certainty to another agent's certainty when it cannot be validated by records from the past.
A by-product of our analysis is an interaction-free detection scheme for the existence of such records.
\end{abstract}

\maketitle

\section{Introduction}
In their recent paper \cite{FraRen}, Frauchiger and Renner propose an interesting Gedan\-ken\-experiment
which can be thought of as an extension of (Deutsch's \cite{Deu} extension of) the classical Wigner's Friend paradox \cite{Wig}. It describes
a protocol involving two labs, $\L$ and $L$, operated by the `friends' $\F$ and $F$,
respectively. A spin $\frac12$-particle $S$, whose state is determined by a quantum coin $R$, is prepared in lab $\L$
and sent to $L$. Subsequently the labs $\L$ and $L$ are measured, in suitably chosen bases, by two `Wigners' $\W$ and $W$.
Frauchiger and Renner argue that, with probability one, $W$ will arrive at contradictory measurement outcomes if one simultaneously accepts
\begin{itemize}
\item[(Q)] quantum mechanics: the rules of quantum mechanics apply to all agents;
\item[(C)] consistency: if agent $A$ has established that ``I am certain that agent $B$ is certain that $x = \xi$ at time $t$'',
then agent $A$ can conclude that ``I am certain that $x = \xi$ at time $t$.''
\end{itemize}
On the basis of the paradox it is argued in \cite{FraRen} that (Q) and (C) are incompatible with
the `single world assumption' (S) that measurement outcomes are unique.
Although there is no explicit reference to the time at which $A$ has established his/her statement,
scrutinising the arguments one finds that this could be a time different from $t$. In fact, the following
more precise version of (C) is actually used: if agent $A$ has established at time $t_0$ that ``I am certain that agent $B$ will be certain that $x = \xi$ at time $t_1\ge t_0$'',
then at time $t_0$ agent $A$ can conclude that ``I am certain that $x = \xi$ at time $t_1$.''

The Frauchiger--Renner paradox has provoked intense discussion and has been analysed from various points of
view \cite{BHW, Bog, Bru, Bub1, Bub2, DFS, FL, He, Kas, Lal, LazHub, Rel, Riz, Sta, Sud, Yan}.
The aim of this article is to analyse and resolve it from the
point of view of Rovelli's relational interpretation of quantum mechanics (RQM)
\cite{Rov} (see also \cite{Fra, LauRov, SmeRov}).
RQM is inspired by general covariance considerations in quantum gravity and
extrapolates the lessons learnt from relativity, which teaches us that
`time' and `space' are observer dependent, to quantum mechanics by
arguing that also the notion of `state' should be considered observer-dependent. In this view,
physics is the study of consistency of the records that different observers give of the observed phenomena. These observers are allowed
to have different accounts as to what their records mean and how they should be interpreted.
Thus RQM aims to minimize ontological claims about the nature of reality, while preserving a notion of objectivity based on the consistency of report of different observers which makes scientific exchange possible and meaningful. An instructive example of
how this works in the case of the EPR paradox has been worked out in \cite{SmeRov}, where the tenets of RQM are eloquently summarised:

\medskip\noindent
{\em ``In RQM, physical reality
is taken to be formed by the individual quantum events
(facts) through which interacting systems (objects) affect
one another. Quantum events are therefore assumed
to exist only in interactions and (this is the central point)
the character of each quantum event is only relative to the
system involved in the interaction. In particular, which
properties any given system $S$ has is only relative to a
physical system $A$ that interacts with S and is affected by
these properties.

If $A$ can keep track of the sequence of her past
interactions with $S$, then $A$ has information about $S$, in the
sense that $S$ and $A$'s degrees of freedom are correlated.
According to RQM, this relational information exhausts
the content of any observer's description of the physical
world.''
}

\medskip\noindent
An essential point of our analysis is therefore to consider the experiment from the perspective
of each agent separately, carefully keeping track of who knows what at what time, and to base all inferences
exclusively on the available information at the given moment from the perspective taken.
The conclusions of our analysis may be stated as follows.

\begin{enumerate}
\item[(1)] The Gedankenexperiment reveals no conflict between (Q) and the following restricted version
of (C):
\begin{quote}
{\em {\rm(C$'$)}: If at time $t_1$ agent $A$ has established that ``I am certain that agent $B$ is certain at time $t_0\le t_1$ that $x = \xi$ at time $t_0$'' and that no information on $x=\xi$ has been erased between times $t_0$ and $t_1$ (i.e., $x=\xi$ has been recorded), then at time $t_1$ agent $A$ can conclude ``I am certain that $x = \xi$ at time $t_0.$'' }
\end{quote}
(cf. Subsection \ref{sec:RFR}.)
\item[(2)] The very existence of records of the measurements by $\F$ and $F$ can lead to different predictions of certain measurement outcomes.
(cf. Subsection \ref{subsec:different}.)
\end{enumerate}

The distinction between (C) and (C$'$) in Conclusion (1) is crucial for the discussion of the Frauchiger--Renner paradox, as
one step in the deduction of the paradox relies on inferences about the past of quantum mechanical systems that keep no records of their history.
Within a different interpretation of quantum mechanics, {a similar objection against this type of inferences} has been put forward in  \cite{Lal}.
Arguing that it is not fair to promote a certainty of one agent to the certainty of another,
RQM rejects (C) (cf. \cite[Table 4]{FraRen}) and dismisses the paradox on those grounds.
Although our Hypothesis (C$'$) is independent of RQM, it is not in contradiction with it, and adopting it instead of (C) makes the Frauchiger--Renner paradox disappear.

Conclusion (2) is a variation on the Elitzur--Vaidman interaction-free detection scheme. It has the interesting additional feature that the records, once created, have no interaction whatsoever with the rest of the system and can nevertheless be detected.

\medskip

The paper is organised as follows. In Section \ref{sec:descr} we describe the Gedankenexperiment
and summarise the main steps in the reasoning in
\cite{FraRen}. In Section \ref{sec:modified} we begin our analysis of the experiment in the framework of RQM
by first leaving out the public announcements of the Wigners. The full experiment is analysed in
Section \ref{sec:RFR}. The main conclusion is that the accounts about the outcome of the experiment given by the various agents from their perspectives
agree at all points with the account of an external observer. In that sense, there is no paradox at all.
In the final Section \ref{sec:discussion} we critically examine the individual steps in the Frauchiger--Renner argument within the RQM formalism. This leads to the conclusions stated above.

\section{Description of the Gedankenexperiment} \label{sec:descr}

Following the notation of \cite{FraRen} we will begin by describing the Gedankenexperiment. It involves four agents: two `Wigners' ($\W$ and $W$) and two `friends' ($\F$ and $F$).
The four agents agree beforehand on a protocol which is repeatedly run until the halting condition specified in the last step is reached.
Each run consists of the following steps.
\begin{enumerate}
 \item[0.] At $t=0$ the following step has been completed: $\F$ prepares a
quantum coin $R$ in the following superposition of the `tail' and `head' states $\ket{t}_R$ and $\ket{h}_R$
$$ \sqrt{\frac23} \ket{t}_R + \sqrt{\frac13} \ket{h}_R$$
and measures it in the $\{\ket{t}_R, \ket{h}_R\}$ basis. If the outcome is `tail', $\F$ sends a spin-$\frac12$ particle $S$ to $F$ in superposition state
 $\frac1{\sqrt 2}\ket{\up}_S + \frac1{\sqrt 2}\ket{\down}_S$; if the outcome is `head', she sends the particle in state $\ket{\down}_S$.
 \item[1.] At $t=1$ the following step has been completed: $F$ measures $S$ in the  $\{\ket{\up}_S,\ket{\down}_S\}$ basis.
 \item[2.] At $t=2$ the following steps have been completed: $\W$ measures the lab $\L = \{R,\F\}$
in an orthonormal basis containing
$\ket{\ovok}_{\L}$ and $\ket{\ovfail}_\L$, where
\begin{equation*}
\begin{aligned} \ket{\ovok}_\L &:= \frac1{\sqrt2}\ket{h}_R \ket{h}_\F - \frac1{\sqrt2}\ket{t}_R\ket{t}_\F \\
\ket{\ovfail}_\L &:= \frac1{\sqrt2}\ket{h}_R \ket{h}_\F + \frac1{\sqrt2}\ket{t}_R\ket{t}_\F,
\end{aligned}
\end{equation*}
and announces the result to everyone.
\item[3.] At $t=3$ the following steps have been completed: $W$ measures the lab $L = \{S,F\}$ in an orthonormal basis containing
$\ket{\ok}_L$ and $\ket{\fail}_L$,
where
\begin{equation*}
\begin{aligned} \ket{\ok}_L &:= \frac1{\sqrt2}\ket{\down}_S \ket{\down}_F - \frac1{\sqrt2}\ket{\up}_S\ket{\up}_F, \\
\ket{\fail}_L &:= \frac1{\sqrt2}\ket{\down}_S \ket{\down}_F + \frac1{\sqrt2}\ket{\up}_S\ket{\up}_F
\end{aligned}
\end{equation*}
and announces the result to everyone.
If $\W$ and $W$ have both announced `ok' (the halting condition) the experiment is halted; otherwise, the protocol is repeated.
 \end{enumerate}
An easy computation, reproduced below, shows that from the perspective of a fifth external observer $C$, in each run the halting condition is reached with probability $\frac1{12}$. It is also shown below that $\W$ and $W$ reach the same conclusion from their perspectives.
Therefore, $\W$ and $W$ agree with $C$ that with probability one the halting condition will be eventually met.

It should be mentioned that \cite{FraRen} only assumes measurements by $\W$ and $W$ in orthonormal bases containing
$\ket{\ovok}_\W$ and $\ket{\ok}_W$, respectively, and that the agents announce `fail' in all cases when their measurement differs from `ok'. Our slightly stronger assumption, where $\ket{\ovfail}_\L$ and $\ket{\fail}_L$ are explicitly added to the bases,
has the advantage of allowing simple explicit computations and does not affect the conclusions of our analysis.

\subsection{The paradox}\label{sec:conundrum}

Of interest is what happens in the final run of the experiment which leads to the halting condition.
In \cite{FraRen} it is argued that if one simultaneously accepts the hypotheses (Q) and (C)
introduced earlier, then the following assertions hold simultaneously:
\begin{itemize}
 \item in each run, with probability $\frac1{12}$ both $W$ and $\W$ measure `ok'.\label{outcome1}
 \item in the final run $W$ is certain to announce `fail'.\label{outcome2}
\end{itemize}
The assertion implies that with probability one the halting condition will be eventually reached. By definition,
in that round $\W$ and $W$ announce `ok'. This contradicts the second assertion.

The argument by Frauchiger and Renner leading to the
second conclusion can be summarised in four main steps as follows:

\medskip

(i) If $\F$  measures `tail' at $t=0$,
she sends the particle in state $\frac1{\sqrt 2}\ket{\up}_S + \frac1{\sqrt 2}\ket{\down}_S$,
and infers that $W$ will announce `fail'.

(ii) If $F$ measures `up' at $t=1$, she is certain that $\F$ must have measured `tail' at $t=0$.
Because of (i) and the consistency hypothesis (C), $F$ therefore is certain
that $W$ will announce `fail'.

(iii) If $\W$ measures `ok' at time $t=2$, he infers that $F$
must have measured the spin to be `up' at $t=1$. Because of (ii), $\W$ is then certain that $F$ is
certain that $W$ will announce `fail'. Hence by (C), $\W$ is certain
that $W$ will announce `fail'.

(iv) If $W$ hears $\W$ announce `ok' at time $t=2$, so by (iii) $W$ is certain that $\W$ is certain
that $W$ will announce `fail'. Hence by (C), $W$ is certain to announce `fail'.

\medskip\noindent
Here we just reproduced the main steps; for the intermediate reasoning leading to
them we refer to the original paper. An analysis from the point of view taken in the present paper
is given in the final section. There, we will argue that steps (ii) and (iii) are problematic
because they involve beliefs about the past in the absence of records
which need revision in the presence of records.

\section{Relational analysis of a modified Gedankenexperiment}\label{sec:modified}

Before analysing the Gedankenexperiment as originally proposed by
Frauchiger and Renner, we introduce a minor modification to their protocol. In the modified protocol, instead of announcing the outcome of his measurement right away, $\W$ keeps it secret until the end of the run. This has the obvious advantage that no assumptions
need to be made as to how the other agents should interpret the information that can be inferred from $\W$'s announcement
and how it affects their descriptions of the experiment. All we assume is hypothesis (Q), according to which the agents may use the rules of quantum mechanics to describe the experiment from their vantage points. Since neither (C) nor (C$'$) is used, this analysis is purely relational and
any conclusions drawn from it
will of course remain valid in the original Frauchiger--Renner scenario, which we consider in the next section.

We will analyse the modified Gedankenexperiment from the points of view of the agents $\F$, $F$, $\W$, $W$, and the external observer $C$ separately. Each of the five agents models the experiment by choosing a tensor product of Hilbert spaces describing the experiment from her/his point of view. For instance, $\F$ models the experiment by using the space $$H_R\otimes H_F\otimes H_S\otimes H_\W\otimes H_W$$ to describe the quantum coin, the friend $F$, the particle $S$, and the Wigners $\W$ and $W$. The description that $\F$ gives of the state at time $t=k$ will be denoted by $$\ket{\Psi}^{\F, t=k}_{RFS\W W}.$$ Similar self-explanatory notation will be used for the other agents's descriptions. Occasionally we will use the notation $$H_\L = H_R\otimes H_\F, \qquad H_L = H_S\otimes H_F$$ to describe the `labs' $\L = \{R,\F\}$ and $L = \{S,F\}$.
The perspective of $C$ serves as a description of the experiment from an external point of view. He is the only agent using
the `full' space $$H_\F\otimes H_R\otimes H_F\otimes H_S\otimes H_\W\otimes H_W$$ involving all four agents plus the coin and the particle. If the paradox were to be experimentally tested, we expect the experimenters to report as $C$ does. The external perspective is included to emphasize that of the contradictory statements of the paradox presented in subsection \ref{sec:conundrum}, the first is correct and the second is false.

From the points of view of $\F$ and $F$ a complete analysis of the experiment is not possible in the mathematical framework outlined above: When $\F$ attempts to describe the joint state of $R$, $S$, $F$, $\W$, and $W$, she will run into difficulties describing $\W$'s
measurement of her lab $\L = \{R,\F\}$, of which she is part. Describing $\L$ from the view of $\F$ amounts to letting $\F$ perform a self-measurement.
A self-measurement scheme would have to describe information erasure from the perspective of $\F$. To our knowledge such a scheme has never been proposed. For a fuller discussion of this issue see \cite{Bre,Lau} and, in the context of the Frauchiger--Renner paradox, \cite{LazHub, Sud}.
Likewise, $F$ will not be able to describe the measurement by $W$ of her lab $L = \{S,F\}$ of which she is part.

\subsection{The point of view of $\F$}
We start by analysing $\F$'s perspective until $\W$ is about to perform his measurement.

We assume that $\F$'s measurement of $R$ results in `tail' (the case of `head' being of no interest here).
After the spin particle has been prepared, at time $t=0$ $\F$ describes the joint state
of $R$, $S$, $F$, $\W$, and $W$ as follows:
 \begin{equation}\label{eq:F0}
     \ket{\Psi}^{\F, t=0}_{RFS\W W}
     = \ket{t}_R\ket{0}_F\Bigl(\frac{1}{\sqrt{2}}\ket{\up}_S + \frac{1}{\sqrt{2}}\ket{\down}_S\Bigr)\ket{0}_\W\ket{0}_{W},
 \end{equation}
where $\ket{0}$ denote ready-to-measure states.
 Now $F$ measures the spin of $S$. According to $\F$ this entangles the states of $S$ and $F$. The resulting joint
state will be described by $\F$ as
 \begin{equation}\label{eq:F1}
\begin{aligned}
     \ket{\Psi}^{\F, t=1}_{RFS\W W}
     & = \ket{t}_R\Bigl(\frac{1}{\sqrt{2}}\ket{\up}_F\ket{\up}_S + \frac{1}{\sqrt{2}}\ket{\down}_F\ket{\down}_S\Bigr)\ket{0}_\W\ket{0}_{W}
\\ & = \ket{t}_R\ket{\fail}_{L}\ket{0}_\W\ket{0}_{W}.
 \end{aligned}\end{equation}
 At this point, $\W$ measures the lab $\L = \{R,\F\}$. The present formalism does not permit $\F$ to make any prediction about $\W$'s
findings because $\F$'s state is not included $\F$'s description of the experiment.

\subsection{The point of view of $F$}
Next we analyse $F$'s perspective until $W$ is about to perform his measurement.

Once $R$ has been prepared and measured by $\F$
and the spin particle has been prepared, at time $t=0$ $F$ describes the joint state
of $R$, $\F$, $S$, $\W$, and $W$ as follows:
\begin{equation}
\begin{aligned}
     \ket{\Psi}^{F, t=0}_{R\F S\W W}
      = \Bigl(\sqrt {\frac23}\ket{t}_R\ket{t}_\F\bigl(\frac1{\sqrt 2}\ket{\up}_S + \frac1{\sqrt 2}\ket{\down}_S\bigr) + \frac1{\sqrt 3}\ket{h}_R\ket{h}_\F\ket{\down}_S\Bigr)\ket{0}_\W\ket{0}_{W}.
\end{aligned}
\end{equation}
Now $F$ measures the spin of $S$.

If she finds $S$ to be `up', the probability of which she evaluated beforehand (at $t=0$) to be $\frac13$, at $t=1$ she describes the new state as
\begin{equation}\label{eq:Fup}
\begin{aligned}
    \ket{\Psi}^{F,t=1}_{R\F S\W W}
    & = \ket{t}_R\ket{t}_\F\ket{\up}_S \ket{0}_\W\ket{0}_{W}
    \\ & = \bigl(\frac1{\sqrt 2}\ket{\ovfail}_\L - \frac1{\sqrt 2}\ket{\ovok}_\L\bigr) \ket{\up}_S
    \ket{0}_\W\ket{0}_{W}.
\end{aligned}
\end{equation}
Consequently she is certain that measuring $R$ and $\F$ at $t=1$ would both result in `up'.
Still assuming she found $S$ to be `up', right after $\W$'s measurement she describes the new state as follows:
\begin{equation}
\begin{aligned}
    \ket{\Psi}^{F,t=2}_{R\F S\W W}
      = \bigl(\frac1{\sqrt 2}\ket{\ovfail}_\L \ket{\ovfail}_\W - \frac1{\sqrt 2}\ket{\ovok}_\L \ket{\ovok}_\W\bigr) \ket{\up}_S \ket{0}_{W}.
\end{aligned}
\end{equation}
She concludes that with probability $\frac13\times \frac12 = \frac16$, $\W$ will have measured `ok'.
Here she is forced to stop her analysis: At $t=3$ her lab, of which she herself is part, is going to be measured by $W$.
Note that if $F$ measures $S$ to be `down', she finds $\ket{\ovfail}_\L$ with probability one.

In conclusion,

\medskip
\begin{center}
\fbox{
    \parbox{11cm}{
Before each run, $F$ evaluates the probability that $\W$ will report `ok' to be $\frac16$.
                 }
     }
\end{center}

\subsection{The point of view of $\W$}\label{subsec:ovW}
The first few steps are similar: after preparing and measuring $R$ and preparing $S$, at $t=0$ $\W$ describes the joint
state of $R$, $\F$, $S$, $F$ and $W$ as
\begin{equation}\label{eq:SovW0}
\begin{aligned}
\ket{\Psi}^{\W, t=0}_{R\F SFW}
=    \Bigl(\frac1{\sqrt 3}\ket{t}_R\ket{t}_\F\bigl(\ket{\up}_S + \ket{\down}_S\bigr)
+ \frac1{\sqrt 3}\ket{h}_R\ket{h}_\F\ket{\down}_S\Bigr)\ket{0}_F\ket{0}_{W}.
\end{aligned}
\end{equation}
Then the spin particle $S$ is measured by $F$. At $t=1$, $\W$ describes the resulting state as
\begin{equation}\label{eq:SovW}
\begin{aligned}
\ &    \ket{\Psi}^{\W, t=1}_{R\F SFW}
  \\ & \quad =    \Bigl(\frac1{\sqrt 3}\ket{t}_R\ket{t}_\F\bigl(\ket{\up}_S\ket{\up}_F + \ket{\down}_S\ket{\down}_F\bigr) + \frac1{\sqrt 3}\ket{h}_R\ket{h}_\F\ket{\down}_S\ket{\down}_F\Bigr)\ket{0}_{W}
\\ & \quad =  \Bigl(\frac1{\sqrt{3}}\bigl(\frac1{\sqrt2}\ket{\ovfail}_\L-\frac1{\sqrt2}\ket{\ovok}_\L\bigr)\bigl(\ket{\up}_S\ket{\up}_F + \ket{\down}_S\ket{\down}_F\bigr)
   \\ & \qquad + \frac1{\sqrt {3}}\bigl(\frac1{\sqrt2}\ket{\ovfail}_\L+\frac1{\sqrt2}\ket{\ovok}_\L\bigr)\ket{\down}_S\ket{\down}_F\Bigr)\ket{0}_W
   \\ & \quad =
\Bigl(\ket{\ovfail}_\L \Bigl(\frac1{\sqrt{6}}\ket{\up}_S\ket{\up}_F
+\sqrt{\frac23}\ket{\down}_S\ket{\down}_F
\Bigr) - \frac1{\sqrt{6}}\ket{\ovok}_\L\ket{\up}_S\ket{\up}_F
 \Bigl)\ket{0}_{W}.
\end{aligned}
\end{equation}
At this point, $\W$ measures the lab $\L = \{R,\F\}$.
We are interested in the scenario that $\W$ measures `ok' which, on the basis of
\eqref{eq:SovW}, he expects to happen with probability $\frac16$.
In that scenario, at $t=2$ he describes the new state as
\begin{equation}\label{eq:ovWok}
    \ket{\Psi}^{\W, t=2}_{R\F SFW}  =  \ket{\ovok}_\L\ket{\up}_S\ket{\up}_F\ket{0}_{W}.
\end{equation}

Rewriting \eqref{eq:ovWok} as
\begin{equation}\label{eq:ovWokW}
   \ket{\ovok}_\L\bigl( \frac1{\sqrt2}\ket{\fail}_L-  \frac1{\sqrt2}\ket{\ok}_L  \bigr)\ket{0}_{W},
\end{equation}
$\W$ infers that if he measures `ok', then $W$ will report `ok' with probability $\frac12$.
Now knowing $W$'s outcome, at $t=3$ he will describe the state by
\begin{equation}
 \ket{\Psi}^{\W, t=3}_{R\F SFW}  =    \ket{\ovok}_\L\bigl( \frac1{\sqrt2}\ket{\fail}_L\ket{\fail}_{W}-  \frac1{\sqrt2}\ket{\ok}_L\ket{\ok}_{W}  \bigr),
\end{equation}

In conclusion,

\medskip
\begin{center}
\fbox{
    \parbox{11cm}{
Before each run, $\W$ evaluates the probability that he will report `ok' to be $\frac16$, and that
if that happens the probability that $W$ will report `ok' is $\frac12$.
Consequently, $\W$ expects that
with probability $\frac1{12}$, both he and $W$ will report `ok'.
                 }
     }
\end{center}

\subsection{The point of view of $W$}
The step $t=0$ is similar as for $\W$ and is skipped. After
the spin particle $S$ has been created by $\F$ and measured by $F$,  $W$ describes the joint state
of $R$, $\F$, $S$, $F$, and $\W$ by
\begin{equation}\label{eq:SFW}
\begin{aligned}
\ &    \ket{\Psi}^{W, t=1}_{R\F SF\W}
  \\ & \quad =    \Bigl(\frac1{\sqrt 3}\ket{t}_R\ket{t}_\F\bigl(\ket{\up}_S\ket{\up}_F + \ket{\down}_S\ket{\down}_F\bigr) + \frac1{\sqrt 3}\ket{h}_R\ket{h}_\F\ket{\down}_S\ket{\down}_F\Bigr)\ket{0}_{\W}
  \\ & \quad =  \Bigl(\frac1{\sqrt{3}}\bigl(\frac1{\sqrt2}\ket{\ovfail}_\L-\frac1{\sqrt2}\ket{\ovok}_\L\bigr)\bigl(\ket{\up}_S\ket{\up}_F + \ket{\down}_S\ket{\down}_F\bigr)
   \\ & \qquad + \frac1{\sqrt {3}}\bigl(\frac1{\sqrt2}\ket{\ovfail}_\L+\frac1{\sqrt2}\ket{\ovok}_\L\bigr)\ket{\down}_S\ket{\down}_F\Bigr)\ket{0}_\W
   \\ & \quad =
\Bigl(\ket{\ovfail}_\L \Bigl(\frac1{\sqrt{6}}\ket{\up}_S\ket{\up}_F
+\sqrt{\frac23}\ket{\down}_S\ket{\down}_F
\Bigr) - \frac1{\sqrt{6}}\ket{\ovok}_\L\ket{\up}_S\ket{\up}_F
 \Bigl)\ket{0}_{\W}.
\end{aligned}
\end{equation}
At this point, $\W$ measures the lab $\L$. From the point of view of $W$ this leads to the description
\begin{equation}\label{eq:measovW}
\begin{aligned}\ &    \ket{\Psi}^{W, t=2}_{R\F SF\W}
  \\ &
 \quad = \ket{\ovfail}_\L \Bigl(\frac1{\sqrt{6}}\ket{\up}_S\ket{\up}_F
+\sqrt{\frac23}\ket{\down}_S\ket{\down}_F
\Bigr) \ket{\ovfail}_{\W}- \frac1{\sqrt{6}}\ket{\ovok}_\L\ket{\up}_S\ket{\up}_F
\ket{\ovok}_{\W}.
\end{aligned}
\end{equation}
The preceding formula can be rewritten as
\begin{equation}\label{eq:W}
 \begin{aligned}
\ & \ket{\ovfail}_\L \Bigl(\frac1{\sqrt{6}}\big(\frac1{\sqrt 2}\ket{\fail}_L - \frac1{\sqrt 2}\ket{\ok}_L\big)
 +\sqrt{\frac23}\big(\frac1{\sqrt 2}\ket{\fail}_L + \frac1{\sqrt 2}\ket{\ok}_L\big)
\Bigr)\ket{\ovfail}_{\W}
\\ & \quad - \frac1{\sqrt{6}}\ket{\ovok}_\L\Big(\frac1{\sqrt 2}\ket{\fail}_L
 - \frac1{\sqrt 2}\ket{\ok}_L\Big) \ket{\ovok}_{\W}
 \\ & = \ket{\ovfail}_\L \Bigl(\sqrt{\frac34} \ket{\fail}_L + \frac1{\sqrt{12}}\ket{\ok}_L
\Bigr)\ket{\ovfail}_{\W}
\\ & \qquad - \frac1{\sqrt{6}}\ket{\ovok}_\L\Big(\frac1{\sqrt 2}\ket{\fail}_L - \frac1{\sqrt 2}\ket{\ok}_L\Big) \ket{\ovok}_{\W} .
 \end{aligned}
 \end{equation}
On the basis of this, $W$ expects with probability $\frac1{12}$ that both he and $\W$ are going to report `ok'.
Also, $W$ expects $\W$ to announce `ok' with probability
$\frac1{12}+\frac1{12}=\frac16$.
Not knowing $\W$'s outcome, upon measuring `ok' $W$ will describe the new state as
\begin{equation}\label{eq:ovW}
 \begin{aligned}
\ket{\Psi}^{W, t=3}_{R\F SF\W} =
\Bigl(\frac1{\sqrt 2} \ket{\ovfail}_\L \ket{\ovfail}_{\W}
+ \frac1{\sqrt{2}}\ket{\ovok}_\L\ket{\ovok}_{\W}\Bigr)\ket{\ok}_L.
 \end{aligned}
 \end{equation}

In conclusion,

\medskip
\begin{center}
\fbox{
    \parbox{11cm}{
Before each run, $W$ evaluates the probability that $\W$ will report `ok' to be $\frac16$, and
the probability that both $\W$ and he will report `ok' to be $\frac1{12}$.
}
     }
\end{center}

\subsection{The external point of view}\label{subsec:externalC}

We now take the point of view of an external observer $C$ (Charlie), who only receives the measurement outcomes
of $\W$ and $W$ but not from $\F$ and $F$ (in this respect $C$ is different from the `super-observer' of \cite{Lal}).

When $F$ measures $S$, at $t=1$ the observer $C$ describes the joint state of $R$, $\F$, $S$, $F$, $\W$, and $W$ as
 \begin{equation}
\begin{aligned}
\ &    \ket{\Psi}^{C, t=1}_{R\F SF\W W}
  \\ & \quad =   \frac{1}{\sqrt{3}}\Bigl(\ket{t}_R\ket{t}_{\F}\bigl(\ket{\uparrow}_S\ket{\uparrow}_F
     + \ket{\downarrow}_S\ket{\downarrow}_F\bigr)
     + \ket{h}_R\ket{h}_{\F}\ket{\downarrow}_S\ket{\downarrow}_F\Bigr)\ket{0}_{\W}\ket{0}_W   .
\end{aligned}
\end{equation}
Next $\W$ measures $\L$.
At $t=2$  $C$ describes the result as
\begin{equation}\label{eq:ovWC}
 \begin{aligned}
\ket{\Psi}^{C, t=2}_{R\F SF\W W} &  =
\Bigl(\ket{\ovfail}_\L \Bigl(\frac1{\sqrt{6}}\ket{\up}_S\ket{\up}_F
+\sqrt{\frac23}\ket{\down}_S\ket{\down}_F
\Bigr)\ket{\ovfail}_{\W}
\\ & \qquad- \frac1{\sqrt{6}}\ket{\ovok}_\L\ket{\up}_S\ket{\up}_F \ket{\ovok}_{\W}
 \Bigl)\ket{0}_{W}.
 \end{aligned}
\end{equation}
On the basis of this, $C$ predicts that $\W$ will announce `ok' with probability $\frac16$.

Then $W$ performs his measurement and, arguing as in \eqref{eq:W}, $C$ updates his description to
\begin{equation}\label{eq:measW}
 \begin{aligned} \ket{\Psi}^{C, t=3}_{R\F SF\W W} & =
\ket{\ovfail}_\L \Bigl(\sqrt{\frac34} \ket{\fail}_L\ket{\fail}_W + \frac1{\sqrt{12}}\ket{\ok}_L\ket{\ok}_W
\Bigr)\ket{\ovfail}_{\W}
\\ & \qquad - \frac1{\sqrt{6}}\ket{\ovok}_\L\Big(\frac1{\sqrt 2}\ket{\fail}_L\ket{\fail}_W - \frac1{\sqrt 2}\ket{\ok}_L\ket{\ok}_W\Big) \ket{\ovok}_{\W}.
 \end{aligned}
 \end{equation}
On the basis of this, $C$ expects that
with probability $\frac1{12}$ both $\W$ and $W$ report `ok'.

We reach the following conclusion:

\medskip
\begin{center}
\fbox{
    \parbox{11cm}{
Before each run, $C$ evaluates the probability that $\W$ will report `ok' to be $\frac16$, and
the probability that both $\W$ and $W$ will report `ok' to be $\frac1{12}$.
                 }
     }
\end{center}

\section{Relational analysis of the Frauchiger--Renner Gedankenexperiment}\label{sec:RFR}

In the original Gedankenexperiment, {\em $\W$ announces the outcome of his measurement before $W$ performs his measurement.}
From this measurement and its subsequent announcement, $\W$ and the other agents are able to deduce various bits of new information.
For instance, when $\W$ measures `ok', in view of \eqref{eq:ovWok} $\W$ is certain that a measurement of $S$ would give `up',
and the other agents are certain at that point that $\W$ is certain that a measurement of $S$ would give `up'.
In contrast to our analysis up to this point, in order to analyse how $\W$'s announcement affects
everyone's analysis we will accept the following principle:

\medskip\noindent
{\bf Hypothesis $({\rm C}')$} (``Shared knowledge of past events''):
If at time $t_1$ agent $A$ has established that ``I am certain that agent $B$ was certain at time $t_0\le t_1$ that $x = \xi$ at time $t_0$ and that
no information on $x=\xi$ has been erased between times $t_0$ and $t_1$ (i.e., $x=\xi$ has been recorded)'', then at time $t_1$ agent $A$ can conclude ``I am certain that $x = \xi$ at time $t_0.$''

\medskip\noindent
This hypothesis allows $A$ to infer new information from announcements made by $B$.
Since $B$ could make his announcement by entering a written record in a notebook which he could pass for inspection to $A$, even from the relational point of view it is reasonable to accept this hypothesis.
It reflects the way we do science and is reasonable on epistemological grounds in that
it offers the possibility to check our beliefs by actually performing a measurement of $X$ at $t=k$
(although admittedly this may change our description of the state and we may not be able to resume the experiment in its original state). \par

 On a philosophical level, (C$'$) can fulfill the role of (C) in the sense that it provides RQM with notion of consistency within the context of quantum mechanical modelling of events. The crucial difference between the two assumptions is that (C$'$) respects the effects of information erasure within quantum mechanics on our reasoning about past events, whereas (C) seems to only function in a classical context.

\medskip\noindent
Let us first take the perspective of $\W$ and assume that he measures `ok'.
He then describes the state of $R$, $\F$, $S$, $F$, $W$ by \eqref{eq:ovWok}, i.e.,
\begin{equation}
    \ket{\Psi}^{\W, t=2}_{R\F SFW}  =  \ket{\ovok}_\L\ket{\up}_S\ket{\up}_F\ket{0}_{W}.
\end{equation}
When $W$ also reports `ok',  $\W$ uses hypothesis $({\rm C}')$ to update his description at $t=3$ to
 \begin{equation}
  \ket{\Psi}^{\W, t=3}_{R\F SFW} = \ket{\ovok}_\L \ket{\ok}_{L}\ket{\ok}_W.
 \end{equation}
From the perspective of $W$,
when $\W$ announces `ok', hypothesis $({\rm C}')$ allows him to update
\eqref{eq:measovW} and \eqref{eq:W} to
\begin{equation}\label{eq:Wbis}
\begin{aligned}  \ket{\Psi}^{W, t=2}_{R\F SF\W}
 & = -\ket{\up}_S\ket{\up}_F  \ket{\ovok}_\L\ket{\ovok}_{\W}
\\ & = -\bigl(\frac1{\sqrt 2}\ket{\fail}_L - \frac1{\sqrt 2}\ket{\ok}_L\big)
 \ket{\ovok}_\L\ket{\ovok}_{\W}.
 \end{aligned}
 \end{equation}
In the event $W$ also measures `ok', he updates his perspective to
\begin{equation}
\begin{aligned} \ket{\Psi}^{W, t=3}_{R\F SF\W} &  =
 \ket{\ok}_L \ket{\ovok}_\L\ket{\ovok}_{\W}.
 \end{aligned}
 \end{equation}

Similarly, upon hearing the announcement `ok' from $\W$, the external observer $C$ uses $({\rm C}')$  to update \eqref{eq:ovWC} to
\begin{equation}\label{eq:WCbis}
 \begin{aligned}
 \ket{\Psi}^{C, t=2}_{R\F SF\W W} &  =
 -\ket{\up}_S\ket{\up}_F \ket{\ovok}_\L \ket{\ovok}_{\W}\ket{0}_{W}
\\ & = -\bigl(\frac1{\sqrt 2}\ket{\fail}_L - \frac1{\sqrt 2}\ket{\ok}_L\big)
\ket{\ovok}_\L\ket{\ovok}_{\W}\ket{0}_{W},
 \end{aligned}
\end{equation}
and, when $W$ also announces `ok', he uses hypothesis $({\rm C}')$ once more to arrive at
\begin{equation}\label{eq:WCbis3}
 \begin{aligned}
 \ket{\Psi}^{C, t=3}_{R\F SF\W W} = \ket{\ok}_L \ket{\ovok}_\L\ket{\ovok}_{\W}\ket{\ok}_{W}.
 \end{aligned}
\end{equation}
Accepting $({\rm C}')$ we reach the following conclusion:

\medskip
\begin{center}
\fbox{
    \parbox{11cm}{
If $\W$ announces `ok', at $t=2$ the agents $\W$, $W$, and $C$ expect
that $W$ will measure `ok' with probability $\frac12$.
                 }
     }
\end{center}

\medskip\noindent
Inspecting various conclusions in this section and the previous one, no Frauchiger--Renner paradox has arisen so far.

\section{Discussion}\label{sec:discussion}

Our analysis of the original Frauchiger--Renner Gedankenexperiment in Section \ref{sec:RFR}, as well as that of the modified
version studied in Section \ref{sec:modified}, reveals that in all three scenarios the involved agents provide
consistent records about the various issues at stake.
Thus the Gedankenexperiment reveals no inconsistency between hypothesis (Q) on the validity of quantum mechanics and hypothesis $({\rm C}')$ on sharing knowledge about the present.

\subsection{Knowing past events}\label{subsec:past} \
Steps (ii) and (iii) employ assertions of the following format:
\begin{quotation}
 {\em If at time $t = l$ agent $A$ knows $p$, then at time $t=l$ agent $A$ is certain that agent $B$ knew $q$ at time $t=k\le l$.}

\end{quotation}
Are assertions of this type to be admitted in scientific reasoning?
If taken literally they belong to the domain of metaphysics, for one cannot check their truth (at least not under (Q)) by time travelling back to $t=k$ and asking $B$ if he really knew. They can only be made into meaningful and testable statements if there are surviving records of $B$'s thoughts, for instance in the form of an entry in a notebook authored by $B$ at $t=k$. The notebook can be accepted by $A$ at time $t=l$
to be a reflection of $B$'s thoughts at time $t=k$. In this process the metaphysical statement ``$A$ is certain that $B$ knew $q$ at time $t=k$'' is reinterpreted as the scientific statement ``$A$ inspects $B$'s notebook at time $t=l$ and finds the entry $q$''.

Frauchiger and Renner employ assertions of the above format in a non-trivial way in at least two places, namely in their derivations of steps (ii) and (iii), through the statements $F^{n:12}$ and $\W^{n:22}$
(cited below) in Table 3 of \cite{FraRen} respectively. If one gives these statements an
operational meaning by equipping the agents with notebooks and re-interpreting the statements accordingly, it turns out that that $F^{n:12}$ can indeed be justified, in the sense that one arrives at the same statement on the basis of the written records, but $\W^{n:22}$ cannot:
a different conclusion is reached on the basis of these records.

Let us now turn to the mathematical details and begin with a look at the first line of step (ii):
\begin{quotation} Statement $F^{n:12}$: \ {\em If $F$ measures `up', then at $t=1$ she is certain that $\F$ must have measured `tail' at $t=0$. }
\end{quotation}
This is statement $F^{n:12}$ in Table 3 of \cite{FraRen} and is of the format discussed above.
We have seen that, upon measuring $S$ to be `up', agent $F$ updates her description of the system to \eqref{eq:Fup}, i.e.,
\begin{equation}\label{eq:Fn12}
\begin{aligned}
    \ket{\Psi}^{F,t=1}_{R\F S\W W}
     = \ket{t}_R\ket{t}_\F\ket{\up}_S \ket{0}_\W\ket{0}_{W}.
\end{aligned}
\end{equation}
Hence she is certain that if {\em she herself} were to measure $R$ {\em at time $t=1$} the outcome would be `tail'.
The statement that, on the basis of this, she is certain that $\F$ must have measured `tail' at $t=0$ is only meaningful in the presence of a surviving record of $\F$'s measurement. As an aside, notice that the particle $S$ itself cannot play the role of this record: for if it could, it would serve also as a record when $F$ measures
`down', but in that case $F$'s description of the system is given by
\begin{equation}\label{eq:ovF1down}
 \begin{aligned}
      \ket{\Psi}^{F,t=1}_{R\F S\W W}
       & = \Bigl(\frac1{\sqrt 2}\ket{t}_R\ket{t}_\F + \frac1{\sqrt 2}\ket{h}_R\ket{h}_\F\Bigr)\ket{\down}_S\ket{0}_\W\ket{0}_{W}
       \\ & = \ket{\ovfail}_\L \ket{\down}_S  \ket{0}_\W\ket{0}_{W},
 \end{aligned}
\end{equation}
and nothing can be recovered from it about $\F$'s measurement of $R$ at $t=0$.

\subsection{Introducing a notebook}\label{subsec:notebook} \
The above discussion motivates us to extend the
protocol by introducing a notebook $\N$ in which $\F$ writes down the result of her measurement at $t=0$. (Likewise,
$F$ needs a notebook $N$ in step (iii); see below). The assertion that $F$ is certain that $\F$ must have measured `tail' at $t=0$ can then be taken to mean that $\F$ is certain that an inspection of the notebook $\N$ {\em at time $t=1$} will reveal
that it contains the entry `tail' (cf. the interpretation of the Gedankenexperiment from the point of
view of QBism in \cite{FraRen}).

In this modified protocol, of which all agents are informed, all agents have to keep track of the state of $\N$ and $N$ as well. They do so by including additional Hilbert spaces $H_\N$ and $H_N$ in their respective descriptions. In the discussion below this is expressed notationally by adding the subscript $\N$ and $N$ respectively. The inclusion of these spaces into all agent's descriptions will be somewhat informally referred to as ``the existence of a notebook'', even if this terminology is somewhat at odds with the tenets of RQM.

Redoing the derivation of \eqref{eq:Fup} with the notebook $\N$ added we arrive at
\begin{equation}\label{eq:notebook_ovF}
\begin{aligned}
    \ket{\Psi}^{F,t=1}_{\N R\F S\W W}
     = \ket{t}_\N\ket{t}_R\ket{t}_\F\ket{\up}_S \ket{0}_\W\ket{0}_{W}.
\end{aligned}
\end{equation}
This provides the justification of $F^{n:12}$.

Turning to the analysis of step (iii),
consider the statement
\begin{quotation} Statement $\W^{n:22}$: \ {\em If $\W$ measures `up', then at $t=2$ he is certain that $F$ must have measured `up' at $t=1$}.
\end{quotation}
This is statement $\W^{n:22}$ in Table 3 of \cite{FraRen} and again it is of the format discussed at the beginning of this subsection.
Providing $\F$ and $F$ with notebooks $\N$ and $N$,
the description at time $t=1$ given by $\W$ becomes
\begin{equation}\label{eq:SovWnote}
\begin{aligned}
 &    \ket{\Psi}^{\W, t=1}_{\N R\F NSFW}
  \\ & \  =    \Bigl(\frac1{\sqrt 3}\ket{t}_\N\ket{t}_R\ket{t}_\F\bigl(\ket{\up}_N\ket{\up}_S\ket{\up}_F + \ket{\down}_N\ket{\down}_S\ket{\down}_F\bigr)
  \\ & \quad \ + \frac1{\sqrt 3}\ket{h}_\N\ket{h}_R\ket{h}_\F\ket{\down}_N\ket{\down}_S\ket{\down}_F\Bigr)\ket{0}_{W}
\\ & \ =  \Bigl(\frac1{\sqrt{3}}\ket{t}_\N\bigl(\frac1{\sqrt2}\ket{\ovfail}_\L-\frac1{\sqrt2}\ket{\ovok}_\L\bigr)\bigl(\ket{\up}_N\ket{\up}_S\ket{\up}_F + \ket{\down}_N\ket{\down}_S\ket{\down}_F\bigr)
   \\ & \quad  + \frac1{\sqrt {3}}\ket{h}_\N\bigl(\frac1{\sqrt2}\ket{\ovfail}_\L+\frac1{\sqrt2}\ket{\ovok}_\L\bigr)\ket{\down}_N\ket{\down}_S\ket{\down}_F\Bigr)\ket{0}_W
   \\ & \ =
\Bigl(\ket{\ovfail}_\L \Bigl(\frac1{\sqrt{6}}\ket{t}_\N\ket{\up}_N\ket{\up}_S\ket{\up}_F
+\frac1{\sqrt 6}\bigl(\ket{t}_\N+\ket{h}_\N\bigr)\ket{\down}_N\ket{\down}_S\ket{\down}_F
\Bigr)
\\ & \quad  + \ket{\ovok}_\L\Bigl(\frac1{\sqrt{6}}\bigl(\ket{h}_\N - \ket{t}_\N \bigr)\ket{\down}_N\ket{\down}_S\ket{\down}_F
 - \frac1{\sqrt{6}}\ket{t}_\N\ket{\up}_N\ket{\up}_S\ket{\up}_F
 \Bigr)\Bigr)\!\ket{0}_{W}.
\end{aligned}
\end{equation}
\vskip-.1cm \noindent If the subsequent measurement of $\L$ gives `ok', at $t=2$ he updates this to
\begin{equation}\label{eq:no-cancel}
\begin{aligned}
   \ &    \ket{\Psi}^{\W, t=2}_{\N R\F NSFW}
  \\ & \ =
 \ket{\ovok}_\L\Bigl(\frac1{\sqrt{3}}\bigl(\ket{h}_\N - \ket{t}_\N \bigr)\ket{\down}_N\ket{\down}_S\ket{\down}_F
 - \frac1{\sqrt{3}}\ket{t}_\N\ket{\up}_N\ket{\up}_S\ket{\up}_F \Bigr)\ket{0}_{W}.
\end{aligned}
\end{equation}
Note the difference with \eqref{eq:ovWok},  i.e.,
\begin{equation}\label{eq:ovWok2}
    \ket{\Psi}^{\W, t=2}_{R\F SFW}  =  \ket{\ovok}_\L\ket{\up}_S\ket{\up}_F\ket{0}_{W},
\end{equation}
where the analogous term to the first term
on the right-hand side of \eqref{eq:no-cancel} disappeared by cancellation.
Due to the presence of $\F$'s notebook $\N$, this cancellation doesn't happen in \eqref{eq:no-cancel} and
$\W$ can no longer be sure
that $F$'s notebook $N$ contains the entry `up'. In this testable sense, he therefore he {\em cannot} claim to ``be certain at $t=2$ that $F$ must have measured `up' at $t=1$''.

This means that $\W^{n:22}$ is false if it is interpreted as a testable statement about the enlarged system that keeps records. We take the point of view that this is the only meaningful interpretation of $\W^{n:22}$.
One could try to maintain the validity of $\W^{n:22}$ as a true statement about the original system without notebooks while at the same time accepting \eqref{eq:no-cancel} as a true statement about the enlarged system with notebooks. But in the absence of records $\W^{n:22}$ becomes a metaphysical statement. The labs $\L$ and $L$ are entangled at the moment of $\W$'s measurement (from $\W$'s perspective, by \eqref{eq:SovW}), and therefore the measurement of $\L$ is in fact a measurement of the joint system comprised of $L$ and $\L$ which erases the memory that $\L$ and $L$ held about the state of $S$ at time $t=1$. At $t=2$, $\W$ has no way of testing the truth of $\W^{n:22}$
by performing measurements on $\L$ and $L$.
All that \eqref{eq:ovWok} and
\eqref{eq:ovWok2} tell us is that $\W$ is certain that if {\em he} were to measure $S$ {\em at time $t=2$} he would find `up'.

We conclude the analysis with a final remark about \eqref{eq:no-cancel}. It is of interest to note that the non-cancellation
disappears if only $F$ is allowed to keep a notebook. Apparently, the existence of recorded information about $R$ makes all the difference, even though statement  $\W^{n:22}$ contains no reference to $R$ whatsoever.
But if only $F$ gets a notebook we run into problems in the earlier analysis of step (ii) where we needed $\F$'s notebook to interpret statement $F^{n:12}$.
One and only one protocol should describe the entire experiment, and therefore we are forced to equip both $\F$ and $F$ with notebooks.

\subsection{Interaction-free detection of records}\label{subsec:different}
Comparison of \eqref{eq:no-cancel} and \eqref{eq:ovWok2} reveals a deeper issue.
What does $\W$ expect to see if he were to enter lab $L$ and measure $S$ at $t=2$? In the original system without notebooks he is certain to find `up', while in the enlarged system with notebooks the probability of finding `up' is down to $\frac13$!
Apparently the mere {\em existence} of the records influences the outcome. Puzzling as this seems, one has to keep in mind that we are comparing statements about different quantum mechanical systems.

This conundrum can be formulated as an interaction-free detection in the spirit of \cite{EliWei}. Suppose that, after agreeing on the Frauchiger--Renner protocol (part of the agreement being that no notebooks will be kept),
$\F$ decides to cheat, takes a hidden notebook from her pocket, and records the outcome secretly.
When $\W$ measures $S$ at $t=2$ he is convinced that \eqref{eq:ovWok2} gives the correct description and therefore expects to see `up' with probability one. Much to his surprise, this happens only in one third of the rounds; in two thirds of the rounds he measures `down'. Then he realises that
the secret use of a notebook by $\F$ would lead to \eqref{eq:no-cancel} and, therefore, that {\em he was able to detect the existence of the secret record inside lab $\L$ by performing measurements inside lab $L$.}

\section{Conclusion}
From the point of view of RQM the Frauchiger–Renner paradox
disappears if one rejects promoting a certainty of one agent's knowledge in the past to the certainty of another in the present agent unless it can be validated by surviving records.
The `paradox’ arises from reasoning on the basis of inferences about the
past using Hypothesis (C), which entails the possibility of reasoning on the basis of past information even when no records are kept of the results. We have shown that if records were to be kept, this would influence the experimental results, even when the agents have no further interaction with the records.
No paradox arises if Hypothesis (C) is replaced by our Hypothesis (C$’$) which limits the use of (C) to scenarios in which no information has been erased.

\medskip\noindent
{\em Acknowledgement} --  The authors thank Lorenzo Catani and Bas Janssens for enlightening discussions.  The  detailed comments and suggestions by the anonymous referees have led to various improvements in the presentation.

\end{document}